\documentclass[letter]{aa} % for the letters 
\usepackage{graphicx}
\usepackage{txfonts}
\usepackage{hyperref}
\usepackage{bm}
\usepackage{pdflscape}
\usepackage{physics}
\newcommand\id{{\mathrm d}}

\newcommand\bu{{\bm u}}

\newcommand\bB{{\bm B}}

\newcommand\Cref{C_{\text{ref}}}
\newcommand\cref{\Cref}

\usepackage{xcolor}

\usepackage{hyphenat}
\begin{document} 

%\title{Subcritical dynamo action in an ultra-hot Jupiter atmosphere: \\ A new scenario for flows and hotspot offset}

\title{Westward hotspot offset explained by subcritical dynamo action in an ultra-hot Jupiter atmosphere}

\titlerunning{Westward hotspot offset explained by subcritical dynamo action in an ultra-hot Jupiter atmosphere}

\author{
        Vincent G. A. Böning\inst{1}
        \and
        Wieland Dietrich\inst{1}
        \and
        Johannes Wicht\inst{1}
%        \and
%        others tbd.\inst{2}
}

\institute{
        {Max-Planck-Institut f\"ur Sonnensystemforschung, Justus-von-Liebig-Weg 3, 37077 G\"ottingen, Germany}\\
        \email{boening@mps.mpg.de}
%        \and
%        {Georg-August-Universit\"at G\"ottingen, Friedrich-Hund-Platz 1, 37077 G\"ottingen, Germany}
        }

\authorrunning{Böning et. al.}

\date{Received ???; Accepted ???}

% \abstract{}{}{}{}{} 
% 5 {} token are mandatory

% \abstract
% % context heading (optional)
% % {} leave it empty if necessary
% {}
% % aims heading (mandatory)
% {}
% % methods heading (mandatory)
% {}
% % results heading (mandatory)
% {}
% % conclusions
% {}

\abstract{Hot Jupiters are tidally-locked Jupiter-sized planets very close to their host star. They have equilibrium temperatures above about 1000 K. Several photometric observations find that the hotspot, the hottest location in the atmosphere, is shifted with respect to the substellar point. Some observations show eastward (prograde) and some show westward hotspot offsets, while hydrodynamic simulations show an eastward offset due to advection by the characteristic eastward mean flow. In particular for ultra-hot Jupiters with equilibrium temperatures above 2000 Kelvin, electromagnetic effects must be considered since the ionization-driven significant electrical conductivity and the subsequent induction of magnetic fields likely result in substantial Lorentz forces. We here provide the first magnetohydrodynamic numerical simulation of an ultra-hot Jupiter atmosphere that fully captures non-linear
electromagnetic induction effects. We simulate an exemplary planet at an equilibrium temperature of about 2400 K. 
We find a new turbulent flow regime, hitherto unknown for hot Jupiters. Its main characteristic is a break-
down of the well-known laminar mean flows. This break-
down is triggered by strong local magnetic fields. We show that the magnetic fields are maintained by a subcritical dynamo process. It is initiated by a sufficiently strong background magnetic field from an assumed deep dynamo region at a realistic amplitude around 2.5 G. Our results show a zero or westward hotspot offset for the dynamo case, depending on atmospheric properties, while the hydrodynamic case has the usual eastward offset. Since our simulation has an eastward mean flow at the equator, radial flows must be important for producing the zero or westward hotspot offset. A subcritical dynamo offers a new scenario for explaining the diversity of observed hotspot offsets. In this scenario, the dynamo has been initiated by sufficiently large fields at some time in the past only for a part of the population.
}%{}{}{}{}


\maketitle

%\tableofcontents

%
%-------------------------------------------------------------------
%-------------------------------------------------------------------

%\newpage

\begin{figure*}
    \centering

    \includegraphics[width=0.98\linewidth]{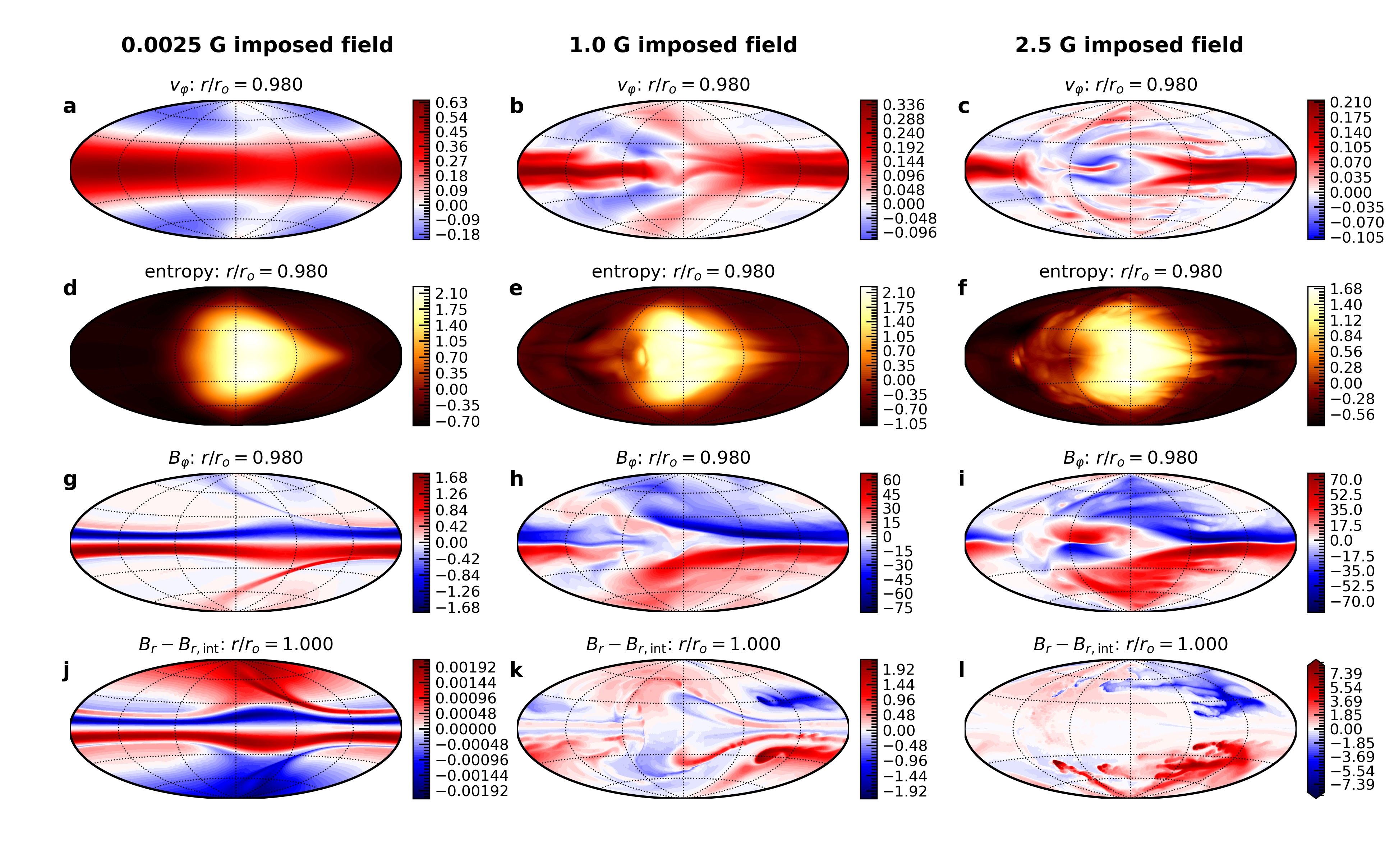}

    \caption{
    Magneto-hydrodynamic simulations of a hot Jupiter atmosphere at equilibrium temperature of 2400 K for three different field strengths of the imposed background dipole (left: 0.025 G; middle: 1.0 G, right: 2.5 G). We show the westward wind speeds as a fraction of the equatorial rotation speed (a,b,c), the entropy profile at a typical radial level (d,e,f), as well as the azimuthal (g,h,i) and radial (j,k,l) magnetic fields at relevant depths in Gauss. For displaying the radial magnetic field, we removed the background dipole component, which reaches its maximum amplitude near the poles of about 65 \% of the imposed field at the bottom of the domain (see panel titles near figure top). The bottom right panel is saturated at one half of the maximum value of $B_r$ at the displayed radial level. 
    See Figure~\ref{figDecay} for the time evolution when the background field is switched off and see Figure~\ref{figUrUThetaBtheta} for plots of $u_r,u_\theta,B_\theta,B_{r,{\rm int}}$ of the snapshots shown in this figure.
    }
    \label{figSubcritical}
\end{figure*}

\section{Introduction}
\label{secIntro}

Since the discovery of the first extra-solar planet, several hundred hot Jupiters (HJs) have been found. These planets orbit their host star in close proximity and 
are therefore locked in rotation; they always face the same side to the star. 
Two observations characterize the potential impact of the atmospheric dynamics on temperature structure: the offset of the brightness maxima from the sub-stellar point 
(hotspot or phase curve offset) and the difference between 
day-side and night-side temperature  \citep[e.g.][]{Parmentier2018,Bell2021,May2022}.
Early observations indicated that hotspots are shifted eastwards (in the prograde direction) from the sub-stellar point \citep[][]{Harrington2006,Cowan2007,Knutson2007} and that the day-to-night-side temperature difference is reduced. Both fit well with the action of a fast eastward directed jet predicted by hydrodynamic theory \citep[e.g.,][]{Showman2011}, shallow general circulation models \citep[e.g.,][]{Showman2002,Heng2011,Rauscher2012,Cho2015}, or hydrodynamic compressible simulations \citep[e.g.,][]{DobbsDixon2008,DobbsDixon2013}. Such a jet would simply advect the temperate structure, and the global dynamics would transport heat from the day side to the night side, reducing the temperature difference \citep[see][for a review]{Showman2020}. 

However, more recent observations \citep[][]{Bell2021,May2022} show that eastward and westward hotspot offsets can occur with no clear dependence on equilibrium temperature or rotation period. No dynamical simulation so far can explain a westward offset, with the exception of a simplified wave model \citep{Hindle2021b}.

The temperatures in the outer atmospheres of HJs can exceed $1500\,$K. This is hot enough to ionize alkali metals and thereby promote electrical conductivities $\sigma$ in the S/m range \citep[e.g., ][]{Kumar2021}. The enormous temperature difference drives fierce zonal winds that 
circumvent the planet. Simulations and observation indicate 
velocities $U$ of the order km/s \citep[e.g.,][]{Showman2020}.

The ability of a flow to generate magnetic fields depends on the non-dimensional magnetic Reynolds number,
\begin{align}
Rm = \mu_0 \sigma U L, \label{eqRmIntro}
\end{align}
where $\mu_0$ is the magnetic permeability and $L$ a haracteristic length scale. Using $\sigma=1\,$S/m, $U=1\,$km/s, and assuming a depth of $L=0.02\, R_J$ for the outer atmosphere, yields a large magnetic Reynolds number of $Rm\approx 10^3$ \citep{Dietrich2022}. The magnetic Reynolds number quantifies the ratio of magnetic field induction to magnetic field dissipation. At such a large value, strong magnetic fields could be generated and impact the atmospheric dynamics \citep{Dietrich2022}.

From theory and earlier work, one would expect two important dynamical regimes that are distinguished by the strength of the Lorentz force \citep[e.g.,][]{Rogers2014,Dietrich2022}. In the weak Lorentz force regime, the magnetic field growth is saturated by magnetic dissipation since the Lorentz force remains too small to change the flows. In the strong Lorentz force regime, however, the Lorentz force provides the dominant saturation mechanism by altering the flows. The Lorentz force can be increased by either larger conductivities (larger temperatures) or stronger background fields. 
Some authors model the impact of the Lorentz force on the flows with a simple linear magnetic drag \citep{Perna2010,Rauscher2013,Beltz2022}. 
Especially for ultra-hot Jupiters with equilibrium temperatures beyond about 1500 Kelvin, stronger magnetic fields will be generated that will change the flow \citep[][]{Dietrich2022}. Here, a fully non-linear treatment of the dynamics and magnetic field generation seems necessary.

At temperatures below about 2400 Kelvin, the conductivity depends strongly on temperature \citep[e.g.,][]{Dietrich2022,Kumar2021}. At larger temperatures, in a range of about 2400 until about 3600 Kelvin, the temperature dependence of the conductivity is only very moderate \citep[][]{Kumar2021,Dietrich2022}, justifying the assumption of a constant conductivity. This happens when the temperature is high enough to fully ionize Alkali metals but not yet high enough to ionize hydrogen. The first nonlinear magnetic models (assuming a constant conductivity) show that the magnetic field may lead to a reversal of the mean zonal winds from eastward to westward \citep{Rogers2014}.The hotspot offset, however, still remains prograde in these models. 
Existing non-linear simulations have probed a regime until about 1900 Kelvin equilibrium temperature \citep{Rogers2014,Rogers2017}. \citet{Rogers2017} show that, when the electrical conductivity varies with background temperature, a self-consistent dynamo can be excited for an ultra-hot Jupiter with a mean temperature of 2300 Kelvin at a pressure of 2 mBar, but they did not report any effect on the hotspot offset. Here, we perform magnetohydrodynamic simulations of an ultra-hot Jupiter atmosphere of an idealized and hypothetical planet at an equilibrium temperature of about 2400 K and analyze the effect on the hotspot offset.

\section{Methods}
\label{secMethods}

We here present numerical simulations of a stably-stratified ultra-hot Jupiter atmosphere, 
building on the anelastic MHD models of flows in a partially stratified domain by \cite{Dietrich2018}, \cite{Gastine2021}, and \cite{Wulff2022} using the MagIC code \citep{Wicht2002,Lago2021}.

We solve the momentum, energy, and induction equations in the anelastic approximation. Details can be found in Appendix~\ref{appEquations}. In the MagIC code, momentum and magnetic fields are decomposed in toroidal and poloidal components, each with a one-dimensional potential. These components can be defined by the requirement that the toroidal field component and the curl of the poloidal component are purely horizontal. The radiative forcing is implemented using a Newtonian cooling term in the energy equation, see Equation~\eqref{eqNewtonian}.

MagIC is a non-dimensional code ruled by the Rayleigh, Ekman, Prandtl, and magnetic Prandtl numbers, as well as non-dimensional numbers that quantify the radial stratification. While it is numerically impossible to simulate the actual parameter regime of a giant planet, we use typical values guided by simulations of the gas giant's interiors, see Table~\ref{tabModelParams} and compare to \citet{Gastine2021}. The stable stratification is chosen in such a way that two characteristic values for hot Jupiters are roughly matched; these are the ratio between buoyancy frequency and rotation frequency, $N/\Omega$, which quantifies the stability of the stratification, and the convective Rossby number, $Ro_c$, which quantifies the relative importance of buoyancy and Coriolis forces. More details are given in Appendix~\ref{appModel}.

To build a magnetic atmospheric model, we use the conductivity values obtained by \cite{Kumar2021}, see Table~\ref{tabModelParams}, and assume a typical ultra-hot equilibrium temperature of $T_{\rm{eq}} = 2400 \, \rm{K}$, which yields a conductivity of $\sigma=0.4\,\rm{S}\,\rm{m}^{-1}$ at a typical density of $\rho_{\rm typ}=7\times 10^{-4} \,\rm{kg}\,\rm{m}^{-3}$ \citep[see App.~\ref{appModel} and][]{Dietrich2022}. We further assume a typical value for rotation frequency of $\Omega=2\times 10^{-5} \,\rm{s}^{-1}$ and a hot Jupiter planetary radius of $R_{HJ}=1.2 R_{J}$, where $R_J$ is Jupiter's radius. Further assuming that the atmosphere has a thickness of $d=0.15 \, R_{HJ}$, using $L=d$ as an estimate for the length scale, and $U=0.2\Omega R_{HJ}$ as an estimate for the flow velocity, this results in a magnetic Reynolds number of $Rm_{HJ}=2140$. The velocity estimate is based on results from non-magnetic simulations. These show that the velocity typically approaches one fifth of the rotation speed of a planet. In other words, the flow Rossby number $Ro=U/(L \Omega)$ is of order 0.2. Using the relation $Rm=\frac{Pm \,Ro}{Ek}=2140$ then allows picking $Pm$ for the simulation, where the Ekman number $Ek$ has been fixed by the hydrodynamic case.

We impose a background axial dipole magnetic field of strength $B_{\rm int}$ at the bottom of the domain, mimicking the action of a dynamo deep in the planet. We perform three different runs with $B_{\rm int} = 0.0025\,\rm{G}$, $1.0\,\rm{G}$, and $2.5\,\rm{G}$ to study the effect of different background magnetic field strengths. For test purposes, we also examine intermediate background field strengths of $B_{\rm int} = 0.025\,\rm{G}$ and $0.25\,\rm{G}$. These field strengths are roughly within the range suggested for hot Jupiters for an internal dynamo. For example, using scaling relations (e.g., \citealp{Dietrich2022}), surface field strengths of Jupiter, $B_J = 5\,\rm{G}$, planetary radii and masses of $R_P/R_J=1.2$ and $M_P/M_J=1.3$, and an age range between 0.1 and 4.5 Gys, one obtains field strengths between 7 and 11 Gauss \citep[compare also to][]{Batygin2010,Yadav2017} as rough upper limits. The scaling shows that the magnetic field strength decreases with age and increases with mass and radius. 

In MagIC, magnetic fields have the unit $\sqrt{\Omega\rho_o /\sigma}$. The square of the non-dimensional field then is the Elsasser number 
\begin{align}
    \Lambda_{\rm int} = \frac{\sigma B_{\rm int}^2}{\rho_o\Omega},
\end{align}
which may serve as a measure for the ratio of Lorentz to Coriolis forces in the Navier-Stokes equation. The internal field values explored here translate to $\Lambda_{\rm int} = 0.0007$, 1, and 6.8, spanning a range where the Lorentz force is expected to have a very small to a very large impact on the dynamics (including the values $\Lambda_{\rm int} = 0.007$ and 0.07 for the test runs). The non-dimensional magnetic field strengths in our runs can be converted to Gauss using a factor of about 0.95.

We integrate the numerical simulations for about a viscous time scale to ensure that a statistically steady state is safely reached. We then test how it further develops if the background dipole field is switched off to test whether the fields are generated self-consistently in a dynamo process. 

Since we chose the parameters of our simulation to reproduce the order of magnitude of the planetary values for $N/\Omega$, the convective Rossby number, and the magnetic Reynolds number, we expect it to reproduce the force balance that rules the dynamics in an ultra-hot Jupiter. 
This is because the ratio $N/\Omega$ and the convective Rossby number $Ro_c$ determine the ratios between buoyancy frequency and rotational timescale, as well as between buoyancy and Coriolis force. See Appendix~\ref{appModel} for further details.

\begin{table}[]
    \centering
    \caption{Assumed parameters for model construction vs. typical ultra-hot Jupiter values.}
    \begin{tabular}{c|c|c|c|c}
         parameter & symb. & unit & simulation &  planet  \\
         \hline
         eq. temp. & $T_{eq}$ & K & 2400 & $ > 2000 $\\
         radius & $R_{HJ}$ & $R_J$& $1.2$ & 0.9- 1.5\\
         rot. freq. & $\Omega$ & $\rm{rad}\,\rm{s}^{-1}$ & $2\times 10^{-5}$ & $2\times 10^{-5}$\\
         typ. density & $\rho_{\rm eq}$ &  $\,\rm{kg} \,\rm{m}^{-3}$ & $7\times10^{-4}$ & $7\times10^{-4}$ \\
         el. conduct. & $\sigma$ & $\rm{S}\,\rm{m}^{-1}$& $0.4$ & $10^{-2}-10 ^1$\\
         shell depth & $d$ & $R_{HJ}$& $0.15$ & $\sim 0.02$\\
         backgr. dipole & $B_{\rm int}$ & G & 0.025 - 2.5 & 0.1-10 \\
         \hline
         Rayleigh & $Ra$ & - & $10^8$ &  $\sim 10^{31}$\\
         Ekman & $Ek$ & - & $10^{-4}$ & $\sim 10^{-15}$\\
         Prandtl & $Pr$ & - & $0.1$ & 0.01 - 1\\
         mag. Prandtl & $Pm$ & - & $0.16$ & $\sim 10^{-12}$\\
         mag. Reynolds & $Rm$ & - & $431^{(a)}$ & $2000^{(b)}$ \\
         Elsasser & $\sqrt{\Lambda_{\rm int}}$ & - & 0.026 - 2.6 & 0.05 - 5\\%\hline
         \hline
         buoy. freq. & $N/\Omega$ & - &  $31.2$ & 100 - 400 \\
         conv. Rossby & $Ro_c$ & - & $3.1$ & 30-1000
    \end{tabular}
    \tablefoot{We give three sets of values, separated by horizontal lines. The first set consists of dimensional paramters assumed for constructing the model for the studied hypothetical hot Jupiter. The second set of values are the corresponding non-dimensional numbers used in the simulation. The third set of values includes the ratio of buoyancy over rotation frequency and the convective Rossby number, which can be used to compare simulations to observations. Further notes: (a) This value is an output from the dynamo simulation. The magnetic Reynolds number is based on the rms. velocity.
    (b) This estimate is based on the conservative assumption that the rms. velocity is about one fifth of the equatorial rotation speed, the same as in the hydrodynamic model, see Appendix~\ref{appModel} and \citet{Dietrich2022}. The difference between the two values for $Rm$ is due to the reduction of the rms. flow velocity in the magnetic compared to the hydrodynamic model.}
    \label{tabModelParams}
\end{table}

\begin{figure*}
    \centering

    \includegraphics[width=0.98\linewidth]{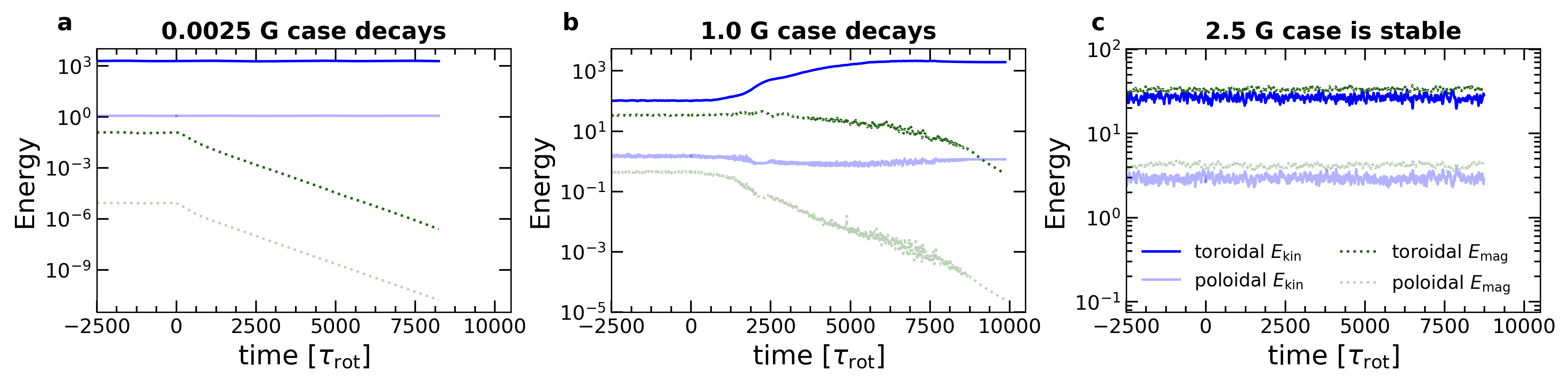}

    \caption{Time evolution of the magnetic and kinetic energies during the stationary state with imposed background field (negative time) and after the background field has been switched off (positive time) for the three cases (a: 0.0025 G; b: 1.0 G, c: 2.5 G). The induced field in the 2.5 G case is stable when the background field is switched off, showing that it is generated by a subcritical dynamo. A decisive criterion for the onset of dynamo action seems that the toroidal magnetic energy exceeds the toroidal kinetic energy.}
    \label{figDecay}
\end{figure*}

\section{Results}
\label{secResults}

Figure~\ref{figSubcritical} shows maps of azimuthal flow (top), entropy (second row), azimuthal magnetic field (third row), and radial magnetic field (bottom) slightly below or at the outer boundary for all three imposed background field strengths. For the weakest imposed field, velocity field and hotspot offset are by eye not distinguishable from the purely hydrodynamic solution. Zonal flows show a typical eastward band in the equatorial region and an eastward hotspot offset. The non-axisymmetric flows are dominated by one horizontal large-scale anticyclonic circulation cell in each hemisphere (Fig.~\ref{figUrUThetaBtheta}). The two cells are centered 30 degrees prograde of the substellar point and about one order of magnitude weaker than the zonal flow. The radial flows are again more than an order of magnitude slower than the horizontal circulation cells and show a large-scale pattern of up and downflows which is slightly shifted and asymmetric with respect to the substellar point. Figure~\ref{figDecay}a shows that the flow is essentially stationary with slow variations on the viscous timescale that amount to roughly 5\% in kinetic energy, which are barely visible in the figure.

The magnetic energy is likewise quasi-stationary and for this weak imposed field much smaller than the kinetic energy (Fig~\ref{figDecay}a). 
Consequently, the Lorentz force is much too weak to affect the wind structure (Fig.~\ref{figSpectra}d and App.~\ref{appTimeEvolEnergiesSpectra}).

Figure~\ref{figSubcritical}j shows that the radial magnetic field is significantly altered away from the imposed axial dipole. Strongly localized stripes are produced around the equatorial plane and connecting the equatorial plane with the poles. A similar pattern also appears in the azimuthal field in Figure~\ref{figSubcritical}g. To understand the origin of this complex magnetic field structure, we analyze the initial time-evolution of a kinematic dynamo simulation started with a purely dipolar background field. The changes to the magnetic field result initially from shearing and advection of the background field. In the growth phase of the field, it progressively assumes a structure that seeks to avoid shear. In other words, the induction term $\nabla \times (\bu \times \bB)$ is minimized since $\bB$ and $\bu$ become almost parallel. In the saturated phase, the remaining induction is balanced by dissipation. For the toroidal field, dissipation plays a larger role than for the poloidal field.

We find that a simple linear scaling $B_{\varphi,\rm max}= Rm\,B_{\rm int}$ explains the maximum toroidal field strength. This scaling is expected for a dissipative balance, which usually is connected to $Rm <1$ \citep{Wicht2019b,Dietrich2022}. In our case, where the induced poloidal field has the same order of magnitude as the weak poloidal background field, the dissipative balance also holds for our large $Rm$ of 431.

We then explore a step-wise increase of the background field to 0.025 G, 0.25 G, and 1.0 G (see Fig.~\ref{figWeakCases} and Fig.~\ref{figSubcritical}). We find that the next significant change happens in the 1.0 G case. Here, the Lorentz force becomes comparable to the Coriolis force for harmonic degrees one and two and even exceeds it at degrees beyond 10, while it is still somewhat smaller at degrees 3 to 9 (Fig.~\ref{figSpectra}e and App.~\ref{appTimeEvolEnergiesSpectra}). The strong Lorentz force significantly alters the flow. The zonal wind underneath and west of the hotspot is substantially suppressed and assumes a wavy pattern (see Fig.~\ref{figSubcritical}b). Non-axisymmetric and time-dependent features now have a magnitude comparable to the axisymmetric zonal flows. Kinetic and magnetic energy vary by a few percent apparently chaotically on timescales of the order of 10 rotation periods (Fig.~\ref{figDecay}b). 
The zonal wind amplitude is reduced by roughly a factor of two. The toroidal and poloidal magnetic fields have been significantly amplified and both are locally much larger than the imposed dipole (see Fig.~\ref{figSubcritical}h,k and Fig.~\ref{figUrUThetaBtheta}h). The field structure becomes localized and very complex.

In the 2.5 G case, the tendency towards a massive change of the flow by the Lorentz force and to introduce new complex space and time dependent features is even stronger compared to the 1.0 G case. Our 2.5 G case and to some extent the 1.0 G case present a completely new dynamical regime of zonal winds on hot Jupiters. In this regime, the influence of magnetic fields is so strong that the zonal flows become unstable and the motions become highly turbulent. The dependence of the fields on time and space is significantly increased in this case. In the 2.5 G case, the Lorentz force is comparable to or large than the Coriolis force at all scales (Fig.~\ref{figSpectra}f and App.~\ref{appTimeEvolEnergiesSpectra}). The toroidal magnetic energy now exceeds the toroidal kinetic energy and the poloidal magnetic energy exceeds the poloidal kinetic energy (Fig.~\ref{figDecay}c). Flow and magnetic field become even more complex. Zonal flows assume a fuzzy structure at nearly all latitudes. A band of prograde wind is still the strongest wind in the domain, but it is completely interrupted at some longitudes (see Fig.~\ref{figSubcritical}c). The radial magnetic field shows a very localized disruptive structure. 
The radial field strength at the surface of the model reaches typical values around 2 G and a maximum value around 10 G (see Fig.~\ref{figSubcritical}l). The azimuthal field $B_\varphi$ reaches typical values of 15-20 Gauss and a maximum value around 80 Gauss (Fig.~\ref{figSubcritical}i).

\begin{figure}
    \centering

    \includegraphics[width=0.98\linewidth]{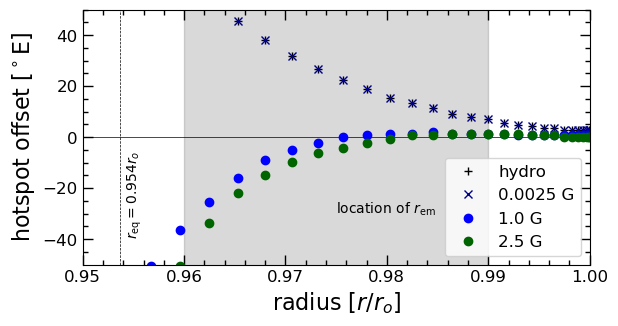}

    \includegraphics[width=0.98\linewidth]{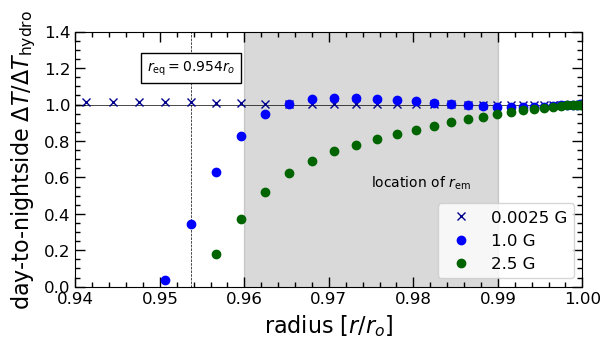}

    \caption{Hotspot offset as a function of depth for the different runs (top) and dayside-to-nightside temperature difference for the magnetic cases relative to the hydrodynamic case (bottom). The possible range of emission layers is indicated in grey. Phase and temperature curves for a typical $r_{\rm em }=0.98 r_o$ are shown in Figures~\ref{figPhasecurves} and \ref{figTempcurves}. The equilibrium temperature is attained at the indicated level of $r_{\rm eq}$.}
    \label{figHotspotshift}
\end{figure}

We observe two important changes in the day-side to night-side heat transfer. Firstly, the changed dynamics in the 1.0 G and 2.5 G cases result in a tendency towards a smaller or even westward hotspot offset, even though the equatorial mean flow remains eastwards (see Fig.~\ref{figHotspotshift}, Fig.~\ref{figSubcritical}, and App.~\ref{appLightcurveHotspotShiftModel}). This indicates that a completely new mechanism controls the hotspot offset, where advective transport by latitudinal or radial flows plays a more important role than by zonal flows. 
Second, the small-scale motions result in a more efficient heat transfer, in particular in the 2.5 G case, which yields a decrease in the day-to-night side temperature difference 
(Fig.~\ref{figHotspotshift} and App.~\ref{appLightcurveHotspotShiftModel}). 
In Figure~\ref{figHotspotshift}, we compare the hotspot offset at different depths in a non-magnetic and our three magnetic cases. A gray background marks the depths from 0.96$r_o$ to $0.99 r_o$, where we expect the strongest contribution to the hotspot observations. For background field strengths of 1 G or larger, the originally positive offset is significantly reduced or even changes sign. A negative offset would be observed if the radiation from radii below 0.98$r_o$ dominated. 
Simplistic models for phase and temperature curves for a typical radial emission level of $r_{\rm em }=0.98 r_o$ are shown in Figures~\ref{figPhasecurves} and \ref{figTempcurves}. A more detailed discussion can be found in Appendix~\ref{appLightcurveHotspotShiftModel}.

To test whether the imposed background fields are strong enough to kick off a subcritical dynamo, we switch off the background field after the simulations have reached a statistically steady state. Subcritical dynamos only work when the magnetic field is strong enough to change the flow in a suitable way. Once the locally produced field has reached a sufficient amplitude, a subcritical dynamo can survive when the imposed field is switched off. 
The time-evolution of the kinetic and magnetic energies of these test runs are shown in Figure~\ref{figDecay}, with the negative part of the time axis indicating the energies of the cases with imposed background field.

When switching off the background field and letting all variables evolve freely, we find that only the magnetic fields from the 2.5 G background case are stable, see Figure~\ref{figDecay}. In this case, the solution is qualitatively unchanged. It has evolved into a self-sustained dynamo which needed the imposed field for a kick start. After reaching a fully developed state, it becomes independent of the background state. This system is therefore a subcritical dynamo.

\section{Discussion}
\label{secDiscussion}

We can only speculate which instabilities set in for a 1.0 G background field and develop to a subcritical dynamo at 2.5 G. From the criterion for the Taylor instability (\citealp{Spruit2002}; Eq. 25 in \citealp{Dietrich2022}), we expect that the onset happens at an azimuthal field strength of the order of $B_\varphi=30$ G. This is locally satisfied in our magnetic simulations for background fields starting at 0.025 G. The Taylor instability sets in preferentially at locations with strong latitudinal gradients in azimuthal magnetic field \citep[Eq.~A.1 in][]{Meduri2024}. The tendency to develop such gradients is already present in the weakest background case. However, a stronger background of at least 1.0 G is apparently required to trigger the onset of the instabilities in our simulations. A more detailed analysis of the instabilities acting in the present simulations may be provided by future work.

A field strength around 2.5 G is of the same order as estimates for the field strength from a deep dynamo on a gas planet \citep[e.g., ][]{Gastine2021}. 
Since the dynamo is subcritical, it is sufficient that the required background field strength is reached at some time in the past. Today's interior dynamo field could actually be weaker. We therefore conclude that almost certainly, there are ultra-hot Jupiters which host an atmospheric dynamo due to the high conductivity.

From our study, we expect that radial fields at the surface reach a typical amplitude of around 1-3 Gauss with local maxima exceeding 10 G. The azimuthal fields deeper in the atmosphere reach amplitudes around 50 - 100 G. A constant conductivity dynamo in ultra-hot Jupiter atmospheres is therefore likely as effective at inducing magnetic fields as a variable conductivity dynamo (where the conductivity is dependent on location, but not on time, see \citealp{Rogers2017}).

Our results are consistent with other studies that employed the anelastic approximation to model hot Jupiter atmospheres \citep{Rogers2014,Rogers2014b,Rogers2017,Rogers2017b}. These authors employed lower conductivities in their constant conductivity models and did not find a dynamo. 
Assuming a temperature of $\sim 2400 \rm{K}$, our conductivity is about 1.5 - 2 times larger than the conductivity in the variable conductivity dynamo obtained by \citet[][at a pressure of 1 bar in their model]{Rogers2017}.

Observations show a diversity of eastward and westward hotspot offsets, particularly around the equilibrium temperature explored here (e.g., \citealp{Bell2021,May2022}). The instabilities discussed here offer a possible explanation. Based on scaling relations \citep[][]{Dietrich2022}, we estimate an order-of-magnitude of 7 to 11 Gauss for the interior field strength of a hot Jupiter. This estimate is, however, rather to be seen as an upper limit \citep{Yadav2017} and field strengths an order of magnitude smaller are also realistic. A planet with such a small interior field will never have experienced the required background field strength of about 2.5 G to kick off the subcritical dynamo and therefore has an eastward hotspot offset. Stronger fields cause a zero or westward hotspot offset if they reach 1 G today or 2.5 G at some time in the past. Since the subcritical dynamo is connected with a change in day-to-nightside temperature difference, this may be used to observationally discriminate between these scenarios.

The subcritical dynamo operates in a highly turbulent flow regime. It is unclear whether such a turbulent flow and temperature pattern can be explained by a simple wave model \citep{Hindle2021b}. For modeling the hotspot offsets of individual ultra-hot Jupiters, a coupling between deep magnetohydrodynamic models with radiative transfer models for observations \citep[e.g.,][]{Zhang2017} is needed in the near future.

\begin{acknowledgements}

We thank U. Christensen, J. Warnecke, and P. Wulff for discussions. VB thanks T. Bell, T. Komacek, E. May, and M. Zhang for detailed explanations of their work. This work was supported by the Deutsche Forschungsgemeinschaft (DFG) in the framework of the priority program SPP 1992 ‘Exploring the Diversity of Extrasolar Planets’. The MAGIC-code is available at an online repository (\url{https://github.com/magic-sph/magic}). This work used NumPy \citep{Oliphant2006,vanderWalt2011}, matplotlib \citep{Hunter2007}, SciPy \citep{Virtanen2020}, and SHTns \citep[][]{Schaeffer2013}.

\end{acknowledgements}

%-------------------------------------------------------------------
% Bibliography
%-------------------------------------------------------------------

%
%\bibliographystyle{aa} % style aa.bst
%\bibliographystyle{astronurl_vb_ads}

%\bibliography{CitaviExport_allCitaviLit}

\begin{thebibliography}{45}
\expandafter\ifx\csname natexlab\endcsname\relax\def\natexlab#1{#1}\fi

\bibitem[{Aubert {et~al.}(2017)Aubert, Gastine, \& Fournier}]{Aubert2017}
Aubert, J., Gastine, T., \& Fournier, A. 2017, Journal of Fluid Mechanics, 813,
  558

\bibitem[{Batygin \& Stevenson(2010)}]{Batygin2010}
Batygin, K. \& Stevenson, D.~J. 2010, Astrophysical Journal Letters, 714, L238

\bibitem[{Bell {et~al.}(2021)Bell, Dang, Cowan, Bean, D{\'e}sert, Fortney,
  Keating, Kempton, Kreidberg, Line, Mansfield, Parmentier, Stevenson, Swain,
  \& Zellem}]{Bell2021}
Bell, T.~J., Dang, L., Cowan, N.~B., {et~al.} 2021, Monthly Notices of the
  Royal Astronomical Society, 504, 3316

\bibitem[{Beltz {et~al.}(2022)Beltz, Rauscher, Roman, \& Guilliat}]{Beltz2022}
Beltz, H., Rauscher, E., Roman, M.~T., \& Guilliat, A. 2022, The Astronomical
  Journal, 163, 35

\bibitem[{Cho {et~al.}(2015)Cho, Polichtchouk, \& Thrastarson}]{Cho2015}
Cho, J. Y.-K., Polichtchouk, I., \& Thrastarson, H.~T. 2015, Monthly Notices of
  the Royal Astronomical Society, 454, 3423

\bibitem[{Cowan {et~al.}(2007)Cowan, Agol, \& Charbonneau}]{Cowan2007}
Cowan, N.~B., Agol, E., \& Charbonneau, D. 2007, Monthly Notices of the Royal
  Astronomical Society, 379, 641

\bibitem[{Dietrich {et~al.}(2022)Dietrich, Kumar, Poser, French, Nettelmann,
  Redmer, \& Wicht}]{Dietrich2022}
Dietrich, W., Kumar, S., Poser, A.~J., {et~al.} 2022, Monthly Notices of the
  Royal Astronomical Society, 517, 3113

\bibitem[{Dietrich \& Wicht(2018)}]{Dietrich2018}
Dietrich, W. \& Wicht, J. 2018, Frontiers in Earth Science, 6

\bibitem[{Dobbs-Dixon \& Agol(2013)}]{DobbsDixon2013}
Dobbs-Dixon, I. \& Agol, E. 2013, MNRAS, 435, 3159

\bibitem[{Dobbs-Dixon \& Lin(2008)}]{DobbsDixon2008}
Dobbs-Dixon, I. \& Lin, D. N.~C. 2008, The Astrophysical Journal Letters, 673,
  513

\bibitem[{Gastine \& Wicht(2021)}]{Gastine2021}
Gastine, T. \& Wicht, J. 2021, Icarus, 368, 114514

\bibitem[{Harrington {et~al.}(2006)Harrington, Hansen, Luszcz, Seager, Deming,
  Menou, Cho, \& Richardson}]{Harrington2006}
Harrington, J., Hansen, B.~M., Luszcz, S.~H., {et~al.} 2006, Science, 314, 623

\bibitem[{Heng {et~al.}(2011)Heng, Menou, \& Phillipps}]{Heng2011}
Heng, K., Menou, K., \& Phillipps, P.~J. 2011, Monthly Notices of the Royal
  Astronomical Society, 413, 2380

\bibitem[{Hindle {et~al.}(2021)Hindle, Bushby, \& Rogers}]{Hindle2021b}
Hindle, A.~W., Bushby, P.~J., \& Rogers, T.~M. 2021, The Astrophysical Journal
  Letters, 922, 176

\bibitem[{Hunter(2007)}]{Hunter2007}
Hunter, J.~D. 2007, Computing in Science {\&} Engineering, 9, 90

\bibitem[{Iro {et~al.}(2005)Iro, B{\'e}zard, \& Guillot}]{Iro2005}
Iro, N., B{\'e}zard, B., \& Guillot, T. 2005, Astronomy {\&} Astrophysics, 436,
  719

\bibitem[{Jones {et~al.}(2011)Jones, Boronski, Brun, Glatzmaier, Gastine,
  Miesch, \& Wicht}]{Jones2011}
Jones, C.~A., Boronski, P., Brun, A.~S., {et~al.} 2011, Icarus, 216, 120

\bibitem[{Knutson {et~al.}(2007)Knutson, Charbonneau, Allen, Fortney, Agol,
  Cowan, Showman, Cooper, \& Megeath}]{Knutson2007}
Knutson, H.~A., Charbonneau, D., Allen, L.~E., {et~al.} 2007, Nature, 447, 183

\bibitem[{Kumar {et~al.}(2021)Kumar, Poser, Sch{\"o}ttler, Kleinschmidt,
  Dietrich, Wicht, French, \& Redmer}]{Kumar2021}
Kumar, S., Poser, A.~J., Sch{\"o}ttler, M., {et~al.} 2021, Physical review. E,
  103, 063203

\bibitem[{Lago {et~al.}(2021)Lago, Gastine, Dannert, Rampp, \&
  Wicht}]{Lago2021}
Lago, R., Gastine, T., Dannert, T., Rampp, M., \& Wicht, J. 2021, Geoscientific
  Model Development, 14, 7477

\bibitem[{May {et~al.}(2021)May, Komacek, Stevenson, Kempton, Bean, Malik, Ih,
  Mansfield, Savel, Deming, Desert, Feng, Fortney, Kataria, Lewis, Morley,
  Rauscher, \& Showman}]{May2021}
May, E.~M., Komacek, T.~D., Stevenson, K.~B., {et~al.} 2021, The Astronomical
  Journal, 162, 158

\bibitem[{May {et~al.}(2022)May, Stevenson, Bean, Bell, Cowan, Dang, Desert,
  Fortney, Keating, Kempton, Komacek, Lewis, Mansfield, Morley, Parmentier,
  Rauscher, Swain, Zellem, \& Showman}]{May2022}
May, E.~M., Stevenson, K.~B., Bean, J.~L., {et~al.} 2022, The Astronomical
  Journal, 163, 256

\bibitem[{Meduri {et~al.}(2024)Meduri, Jouve, \& Ligni{\`e}res}]{Meduri2024}
Meduri, D.~G., Jouve, L., \& Ligni{\`e}res, F. 2024, Astronomy {\&}
  Astrophysics, 683, A12

\bibitem[{Oliphant(2006)}]{Oliphant2006}
Oliphant, T.~E. 2006, A guide to NumPy, Vol.~1 ({Trelgol Publishing USA})

\bibitem[{Parmentier \& Crossfield(2018)}]{Parmentier2018}
Parmentier, V. \& Crossfield, I. J.~M. 2018, in Handbook of Exoplanets, ed.
  H.-J. Deeg \& J.~A. Belmonte, Springer eBook Collection (Cham: Springer),
  1419--1440

\bibitem[{Perna {et~al.}(2010)Perna, Menou, \& Rauscher}]{Perna2010}
Perna, R., Menou, K., \& Rauscher, E. 2010, The Astrophysical Journal Letters,
  719, 1421

\bibitem[{Rauscher \& Menou(2012)}]{Rauscher2012}
Rauscher, E. \& Menou, K. 2012, The Astrophysical Journal Letters, 745, 78

\bibitem[{Rauscher \& Menou(2013)}]{Rauscher2013}
Rauscher, E. \& Menou, K. 2013, The Astrophysical Journal Letters, 764, 103

\bibitem[{Rogers(2017)}]{Rogers2017b}
Rogers, T.~M. 2017, Nature Astronomy, 1, 0131

\bibitem[{Rogers \& Komacek(2014)}]{Rogers2014}
Rogers, T.~M. \& Komacek, T.~D. 2014, The Astrophysical Journal Letters, 794,
  132

\bibitem[{Rogers \& McElwaine(2017)}]{Rogers2017}
Rogers, T.~M. \& McElwaine, J.~N. 2017, Astrophysical Journal Letters, 841, L26

\bibitem[{Rogers \& Showman(2014)}]{Rogers2014b}
Rogers, T.~M. \& Showman, A.~P. 2014, Astrophysical Journal Letters, 782, L4

\bibitem[{Schaeffer(2013)}]{Schaeffer2013}
Schaeffer, N. 2013, Geochemistry, Geophysics, Geosystems, 14, 751

\bibitem[{Schwaiger {et~al.}(2019)Schwaiger, Gastine, \&
  Aubert}]{Schwaiger2019}
Schwaiger, T., Gastine, T., \& Aubert, J. 2019, Geophysical Journal of the
  Royal Astronomical Society, 219, S101

\bibitem[{Showman \& Guillot(2002)}]{Showman2002}
Showman, A.~P. \& Guillot, T. 2002, Astronomy {\&} Astrophysics, 385, 166

\bibitem[{Showman \& Polvani(2011)}]{Showman2011}
Showman, A.~P. \& Polvani, L.~M. 2011, The Astrophysical Journal, 738, 71

\bibitem[{Showman {et~al.}(2020)Showman, Tan, \& Parmentier}]{Showman2020}
Showman, A.~P., Tan, X., \& Parmentier, V. 2020, Space Science Reviews, 216

\bibitem[{Spruit(2002)}]{Spruit2002}
Spruit, H.~C. 2002, Astronomy {\&} Astrophysics, 381, 923

\bibitem[{{van der Walt} {et~al.}(2011){van der Walt}, Colbert, \&
  Varoquaux}]{vanderWalt2011}
{van der Walt}, S., Colbert, S.~C., \& Varoquaux, G. 2011, Computing in Science
  {\&} Engineering, 13, 22

\bibitem[{Virtanen {et~al.}(2020)Virtanen, Gommers, Oliphant, Haberland, Reddy,
  Cournapeau, Burovski, Peterson, Weckesser, Bright, {van der Walt}, Brett,
  Wilson, Millman, Mayorov, Nelson, Jones, Kern, Larson, Carey, Polat, Feng,
  Moore, VanderPlas, Laxalde, Perktold, Cimrman, Henriksen, Quintero, Harris,
  Archibald, Ribeiro, Pedregosa, \& {van Mulbregt}}]{Virtanen2020}
Virtanen, P., Gommers, R., Oliphant, T.~E., {et~al.} 2020, Nature methods, 17,
  261

\bibitem[{Wicht(2002)}]{Wicht2002}
Wicht, J. 2002, Physics of the Earth and Planetary Interiors, 132, 281

\bibitem[{Wicht {et~al.}(2019)Wicht, Gastine, \& Duarte}]{Wicht2019b}
Wicht, J., Gastine, T., \& Duarte, L. D.~V. 2019, Journal of Geophysical
  Research: Planets, 124, 837

\bibitem[{Wulff {et~al.}(2022)Wulff, Dietrich, Christensen, \&
  Wicht}]{Wulff2022}
Wulff, P.~N., Dietrich, W., Christensen, U.~R., \& Wicht, J. 2022, MNRAS, 517,
  5584

\bibitem[{Yadav \& Thorngren(2017)}]{Yadav2017}
Yadav, R.~K. \& Thorngren, D.~P. 2017, Astrophysical Journal Letters, 849, L12

\bibitem[{Zhang {et~al.}(2017)Zhang, Kempton, \& Rauscher}]{Zhang2017}
Zhang, J., Kempton, E. M.-R., \& Rauscher, E. 2017, The Astrophysical Journal
  Letters, 851, 84

\end{thebibliography}

\appendix

\section{Governing equations and numerical method}
\label{appEquations}

Following \citet{Dietrich2018} for the setup in a stably stratified shell, we solve the momentum, energy, and induction equations in the anelastic approximation, which we here write in terms of the rotational timescale, $\tau_{\rm rot} = \Omega^{-1}$,
\begin{align}
\frac{1}{Ek^2}\left(\dfrac{\partial \vec{u}}{\partial t}+\vec{u}\cdot\vec{\nabla}\vec{u} \right)
&= -\vec{\nabla}\left({\dfrac{p}{\tilde{\rho}}}\right) - \dfrac{2}{Ek}\,\vec{e_z}\times\vec{u}
+ \dfrac{Ra}{Pr}\tilde{g} \,S\,\vec{e_r} \nonumber \\
&\quad 
+ \dfrac{1}{Pm \,Ek\,\tilde{\rho}}\left(\vec{\nabla}\times \vec{B}
\right)\times \vec{B}
+ \dfrac{1}{\tilde{\rho}} \vec{\nabla}\cdot \mathcal{S}, \label{eqNSE} \\
\frac{\tilde{\rho}\tilde{T}}{Ek}\left(\dfrac{\partial S}{\partial t} +
\vec{u}\cdot\vec{\nabla} S + u_r \dv{\tilde S}{r}\right) &=
\dfrac{1}{Pr}\vec{\nabla}\cdot\left(\tilde{\rho}\tilde{T}\vec{\nabla} S\right) + \dfrac{Pr\,Di}{Ra}\Phi_\nu \nonumber \\
& \quad +
\dfrac{Pr\,Di}{Pm^2\,Ek\,Ra}\left(\vec{\nabla}
\times\vec{B}\right)^2 + \tilde{\rho}\tilde{T}F_{\text{Newt.}} \label{eqs} \\
\frac{1}{Ek}\dfrac{\partial \vec{B}}{\partial t} &= \vec{\nabla} \times \left( \vec{u} \times \vec{B} \right) -\dfrac{1}{Pm}\vec{\nabla}\times\left(\vec{\nabla}\times\vec{B}\right)\label{eqInduction},
\end{align}
where bold symbols are vectors, $\bu$ is velocity, $\bB$ is magnetic field, $t$ is time, $p$ is pressure, $\rho$ is density, $\vec{e_i}$ is the unit vector in direction $i$, $g$ is gravity, $S$ denotes entropy perturbations, $T$ is temperature, and a tilde denotes background quantities that depend only on radius $r$. The computational domain is a spherical shell with inner radius $r_i$ and outer radius $r_o$. Equations~\eqref{eqNSE}-\eqref{eqInduction} are  complemented by the continuity equation and selenoidal condition,
\begin{align}
\vec{\nabla}\cdot\tilde{\rho}\vec{u}&=0,   \quad\label{eqMassCons}  \vec{\nabla}\cdot\vec{B}=0.
\end{align}

\begin{figure}
    \centering

    \includegraphics[width=0.98\linewidth]{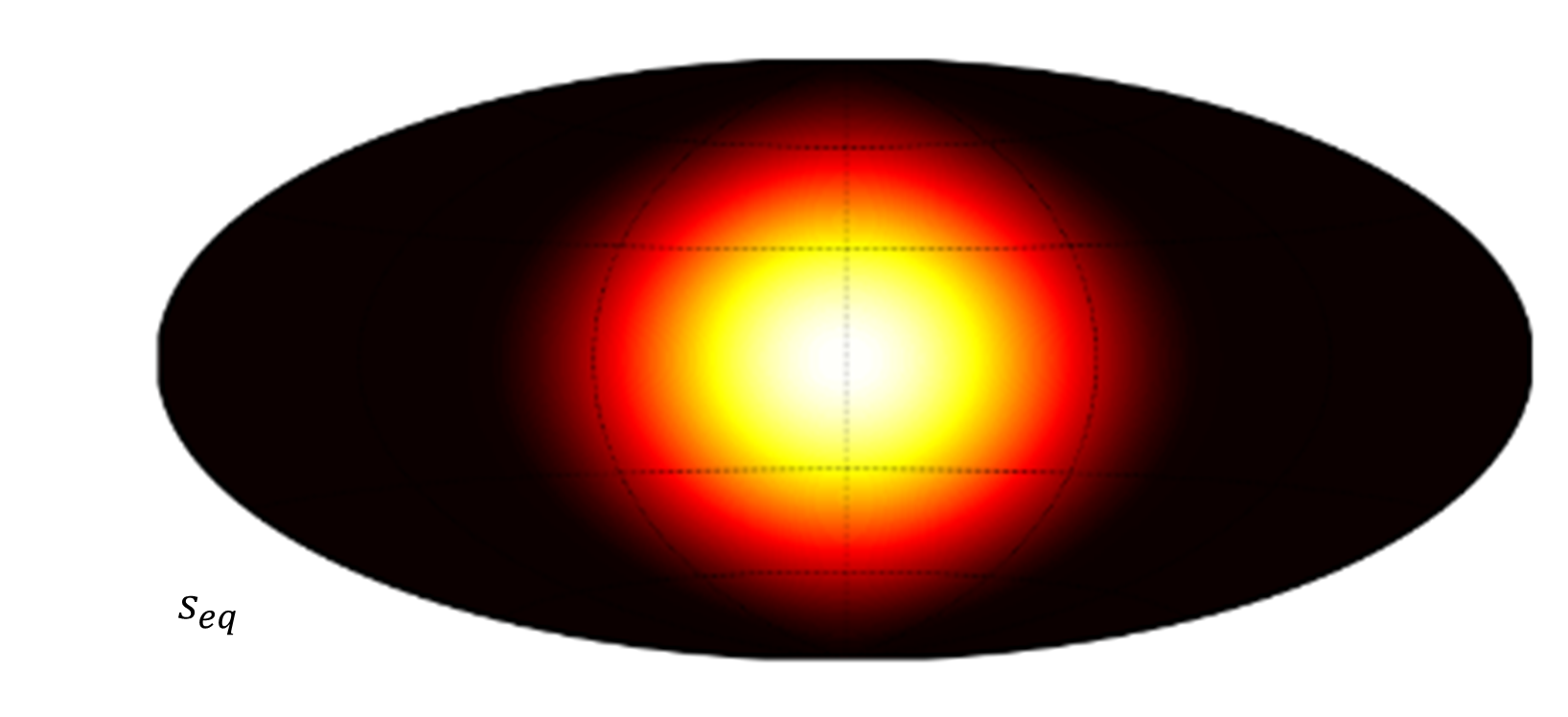}

    \caption{Stellar irradiation pattern used for forcing the entropy profile with $S_{eq}$ in the Newtonian cooling scheme, see Equations~\eqref{eqNewtonian} and~\eqref{eqSeq}.}
    \label{figDriving}
\end{figure}

The rate-of-strain tensor and viscous dissipation into heat are given by (in Cartesian coordinates)
\begin{align}
\mathcal{S}_{ij}  &= 2\tilde\rho e_{ij} - \dfrac{2\tilde\rho}{3}\,(\vec{\nabla}\cdot\vec{u}) \, \delta_{ij}, \label{eqSij} \\
e_{ij} &=\dfrac{1}{2}\left(\dfrac{\partial u_i}{\partial x_j}+\dfrac{\partial
 u_j}{\partial x_i}\right),  \\
\Phi_\nu &= 2\tilde\rho\left[e_{ij}e_{ji}-\dfrac{1}{3}\left(\vec{\nabla}\cdot\vec{u}\right)^2\right].
\end{align}

The non-dimensional formulation of the problem necessitates a comparison to physical numbers, which we provide here. The non-dimensional Rayleigh, Prandtl, magnetic Prandtl, and Ekman numbers are
\begin{align}
Ra &= \frac{\alpha_o g_o T_o d^4 }{c_p \kappa_o \nu_o} \, \Big\vert \dv{\tilde S}{r} \Big\vert_o,  \\
Pr &= \frac{\nu_o}{\kappa_o},\\
Pm &= \frac{\nu_o}{\lambda_o}, \\
Ek &= \frac{\nu_o}{\Omega d^2}.
\end{align}
Here, $\Omega$ is the rotation rate, $c_p$ is the specific heat capacity at constant pressure, $\lambda_o=\frac{1}{\mu_0 \sigma_o}$ is magnetic diffusivity, $\mu_0$ is magnetic permeability,  and the quantities $\sigma_o$, $\alpha_o$, $\nu_o$, $\kappa_o$, $g_o$, $\Big\vert \dv*{\tilde S}{r} \Big\vert_o$
are reference values for electrical conductivity, thermal expansion coefficient, viscosity, thermal conductivity, gravity, and entropy gradient at the outer boundary of the shell, which has a thickness $d=r_o-r_i$.

In the entropy equation \eqref{eqs}, the Newtonian cooling term
\begin{align}
F_{\text{Newt.}} &= \frac{S_{\rm eq} - S}{\tau_{\rm rad}(r)}, \label{eqNewtonian}
\end{align}
includes an equilibrium entropy profile $S_{\rm eq}$ driven by stellar irradiation (see Fig.~\ref{figDriving}),
\begin{align}
    S_{\rm eq}(\theta,\varphi) &= 
\begin{cases}
    S_{\rm min},& \text{if } \frac{1}{2}\pi \leq \varphi \leq \frac{3}{2} \pi\\
    S_{\rm max} \frac{\cos(\alpha) +1}{2}- S_{\rm min},              & \text{otherwise}
\end{cases} \label{eqSeq}
\end{align}
where $\alpha(\theta,\varphi)$ is the angle between the vector $\vec{e_r}(\theta,\phi)$ and the vector pointing towards the substellar point, $\vec{e_r}(\pi/2,0)$. In our model, the minimum and maximum values, $S_{\rm min}$ and $S_{\rm max}$, are related by $S_{\rm min} = - S_{\rm max} / 5$ to guarantee energy conservation.

The radiative timescale $\tau_{\rm rad}(r)$ rapidly increases in the planetary interior,
\begin{align}
\tau_{\rm rad}(r) &= \tau_0 \, (\tilde  \rho \tilde T)^{\tau_{\rm slope}} \label{eqtaurad}.
\end{align}
The Newtonian forcing is implemented in such a way that the simulation domain begins at a radial level somewhat above the level where the dynamics become dependent on radius and are dynamically governed by other effects than the radiative equilibrium. We therefore assume the top of our spherical shell to be at a level where the radiative timescale is about one tenth of the rotational timescale, that is at $\tau_0=0.1$. This choice means that at the outer shell boundary, the radiative timescale safely is the shortest timescale in the system, and becomes of a similar magnitude as the rotational timescale in deeper layers. In the topmost layers, the entropy profile is therefore forced to the equilibrium entropy profile. From the top, the radiative timescale increases according to Equation~\eqref{eqtaurad}, using $\tau_{\rm slope}=2$, consistent with \citet[][Fig. 4]{Iro2005}.

The numerical method used here is in principle very similar to the anelastic formulation used by \citet{Rogers2014}. We solve for an energy-conserving formulation of the equations using the MagIC code (version 5.8, see \citealp{Wicht2002,Lago2021}), which has been benchmarked against other anelastic codes \citep{Jones2011} and extensively used for simulating shells with a partly stable stratification \citep[e.g.,][]{Dietrich2018,Gastine2021,Wulff2022}. As a pseudo-spectral code, MagIC is based on a spherical harmonic transform in the horizontal direction and on a decomposition in Chebychev polynomials in the radial direction. More details can be found in \citet{Wicht2002} and \citet{Gastine2021}.

\section{Atmospheric model}
\label{appModel}

\begin{figure*}
    \centering

    \includegraphics[width=0.98\linewidth]{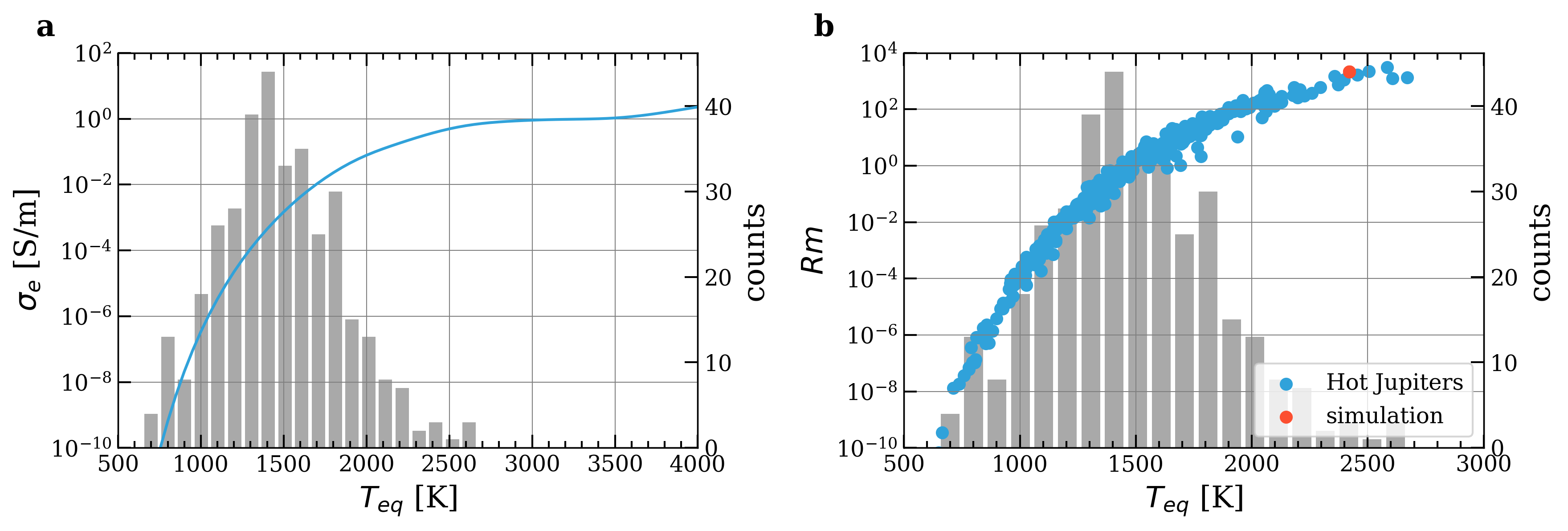}

    \caption{Electrical conductivity (a) at a density of $\rho_{\rm typ}=7\times 10^{-4} \rm{kg}\,\rm{m}^{-3}$ from \citet{Kumar2021} and estimated magnetic Reynolds number (b) as a function of equilibrium temperature. The histograms and blue dots refer to the HJ population studied in \citet{Dietrich2022}.}
    \label{figModel}
\end{figure*}

For the atmospheric model, we assume an equilibrium temperature of $T_{\rm{eq}}=2400\,\rm{K}$, for which the electrical conductivity profiles of \citet{Kumar2021} yield a conductivity of $\sigma=0.4\,\rm{S}\,\rm{m}^{-1}$ at a typical density of $\rho_{\rm typ}=7\times 10^{-4} \,\rm{kg}\,\rm{m}^{-3}$ \citep{Dietrich2022}. We further assume a typical value for rotation frequency of $\Omega=2\times 10^{-5} \,\rm{s}^{-1}$ and a hot Jupiter planetary radius of $R_{HJ}=1.2 R_{J}$, where $R_J$ is Jupiter's radius. We illustrate the choice of electrical conductivity compared to a statistical sample of observed hot Jupiters in Figure~\ref{figModel}.

We further assume that the considered part of the atmosphere has a thickness of $d=0.15 \, R_{HJ}$. This rather large assumption is due to two reasons. First, the pressure scale height in a hot Jupiter atmosphere is expected to be rather large. 
Second, numerical studies of shallower depths than explored here are extremely demanding and require a much higher resolution not obtainable with today's resources. We use a polytropic index of the atmosphere of $m = 2$. The stratification in the shell can further be characterized by the dissipation number,
\begin{align}
    Di &= \frac{\alpha_o g_o d}{c_p} 
    = 3, 
\end{align}
which results in a density variation of $N_\rho = \ln(\tilde{\rho}_i/\tilde{\rho}_o) \approx 3$ density scale heights across the shell. 
We use radial profiles of thermal diffusivity from \citet{Dietrich2018} and a small relative deviation from the adiabat, $\epsilon_s=10^{-4}$.

To roughly estimate where the location of the radiative surface of the planet would be in our model, we first determine the radius $r_{\rm eq}$, where the radiative and rotational timescales are equal, that is $\tau_{\rm rad } = \tau_\Omega$, where $\tau_\Omega=2\pi$. We find $r_{\rm eq}= 0.954 r_o$. At this level, we assume the background temperature to be equal to the equilibrium temperature,
\begin{align}
    \tilde T(r_{\rm eq}) T_o &= T_{\rm eq}
\end{align}
and the background density to be equal to the assumed typical density $\rho_{\rm typ}$,
\begin{align}
    \tilde \rho(r_{\rm eq}) \, \rho_o &= \rho_{\rm typ}
\end{align}
Following our background stratification, we then have $\rho_o = 3.2\times 10^{-4} \,\rm{kg}\,\rm{m}^{-3}$ and  $T_o=1600 \, \rm{K}$. The values of $\nu=\nu_o$ and $\sigma$ are constant across the shell and are therefore chosen at a temperature of 2400 K and density $\rho_{\rm typ}$.

Estimates of the maximum value $S_{\rm max}$ of $S_{\rm eq}$ vary between about 0.3 and 5, depending on stellar and planetary properties. We here choose $S_{\rm max} = 3.33$ for an initial model of a rather hot planet. Details will have to be modeled for each individual planet in subsequent work.

We further choose the following values for the Rayleigh, Prandtl and Ekman numbers, $Ra = 10^8$, $Pr = 0.1$, and $Ek = 10^{-4}$. For the radial entropy gradient of the stable stratification, we choose
\begin{align}
A &= \dv{\tilde S}{r} = 100.
\end{align}
In Table~\ref{tabModelParams}, we summarize the choice of model parameters and compare to a typical hot Jupiter. The chosen parameters result from a compromise between realism and numerical feasability. However, most importantly for comparing with planetary values, we replicate realistic values for the ratio of buoyancy frequency over rotation frequency,
\begin{align}
N/\Omega &= \sqrt{Ra A / Pr} Ek = 31.6,
\end{align}
which is a measure of the ratio between buoyancy and rotation timescales, and the convective Rossby number,
\begin{align}
Ro_c &= \sqrt{Ra / Pr} Ek = 3.16,
\end{align}
which is a measure for the relative importance of buoyancy and Coriolis forces. 
The fact that both numbers match the order of magnitude expected for hot Jupiters indicates that we obtain the relevant force balance in the hydrodynamic simulation.

The relevance of magnetic fields and thus of the Lorentz force is represented by the magnetic Reynolds number \citep{Dietrich2022},
\begin{align}
    Rm &= \mu_0 \sigma U L,
\end{align}
which measures the relative importance of magnetic induction over magnetic diffusion. Here, $U$ is a typical rms. flow velocity, $L$ a typical length scale, $\sigma$ is the electrical conductivity, $\mu_0$ is the magnetic permeability, and $\lambda=1/(\mu_o\sigma)$ is the magnetic diffusivity. Using $U=0.2\Omega R_{HJ}$, which is an estimate we obtain from the hydrodynamic simulation, and $L=d$ as an order-of-magnitude estimate for the lengthscale, we estimate a typical magnetic Reynolds number of $Rm_{HJ}=2140$ for a 2400 Kelvin hot Jupiter. Based on these assumptions, the magnetic Reynolds number can easily be estiamted for a large range of hot Jupiters, which we show in Figure~\ref{figModel}b. The values that we estimate here are a factor of about five lower than those in \citet{Dietrich2022} because these authors used an assumption of $U=\Omega R_{HJ}$ without the empirical factor of 0.2 that we infer here from the hydrodynamic simulation.

In three-dimensional hydrodynamic simulations of planets, realistic parameters, especially for the gas viscosity, cannot be simulated with numerical ressources available today. We therefore first simulate the hydrodynamic case, which results in a Reynolds number, $Re = UL/\nu$, of $Re_{\text{hydro}}=13400$. Using $U=0.2 r_o \,Ek^{-1}$ as the nondimensional rms. flow velocity, we then obtain an estimate for $Pm$ via
\begin{align}
    Pm = \frac{Rm_{HJ}}{Re_{\text{hydro}}} = 0.16.
\end{align}
Alternatively, assuming a turbulent viscosity of $\nu_o=3.2 \times 10^5\,\rm{m}^2\,\rm{s}^{-1}$, which is consistent with the assumed Ekman number and the assumed values for shell thickness and rotation frequency, we obtain the same value for $Pm$ using the electrical conductivity from \citet{Kumar2021} and
\begin{align}
    Pm &= \frac{\nu_o}{\lambda} = \nu_o \, \mu_0 \sigma.
\end{align}

Although estimates of the magnetic Reynolds number are in general very approximate, the magnetic Reynolds number is the most relevant quantity that describes the physics at play in the modeled system. This is because it quantifies the ability of the system to induce and to amplify magnetic fields. The equilibrium temperature and hence the electrical conductivity are also to be considered approximate estimates in our model and are in fact less meaningful quantities than the magnetic Reynolds number. This is to be taken into account when comparing our results to other simulations and to observations.

The values that we chose for the Ekman and magnetic Prandtl numbers are in fact much too large compared to the realistic values, while the Rayleigh number is much lower than the realistic value (see Tab.~\ref{tabModelParams}). Realistic parameters would require a numerical resolution that is computationally by far not affordable, the same as in all simulations of astrophysical fluids. The fact that still a realistic magnetic Reynolds number can be attained can be connected to the relation
\begin{align}
    Rm = Ro \frac{Pm}{Ek}
\end{align}
and to the fact that a realistic Rossby number $Ro=U/(\Omega \,d)$ and a realistic ratio $Pm/Ek$ can be attained. In dimensional terms, this corresponds to simulating realistic quantities except for the viscosity and the entropy gradient at the outer and inner boundaries.

As boundary conditions for entropy, we use a fixed radial gradient, which is fixed to a non-dimensional value of -1 at both boundaries. For velocity, we use stress-free inner and outer boundaries and for magnetic field, we use insulating inner and outer boundaries.

For this setup, we perform magneto-hydrodynamic runs for different background magnetic field strengths. The magnetic field strength in our models is non-dimensionalized in terms of (the square root of the) Elsasser number,
\begin{align}
    B &= \frac{\hat B}{(\mu_0 \lambda \rho_o\Omega)^{1/2}},
\end{align}
where dimensional values of $\mu_0,\rho_0,\Omega$ can be used to obtain a dimensional estimate for the magnetic field, $\hat B$, from the nondimensional $B$ by multiplying with the factor $(\mu_0 \lambda \rho_o\Omega)^{1/2}=0.95$.

Finally, we used hyperdiffusion to damp away some spurious oscillations at high spatial frequency, a phenomenon that apparently has been obsevered also by \citet{Rogers2014}. The hyperdiffusion is implemented as a multiplication of the diffusion operator in spherical harmonic space by the factor
\begin{align}
    d(\ell)=1+D\left[\frac{\ell+1-\ell_{hd}}{\ell_{\rm{max}}+1-\ell_{hd}} \right]^{\beta}
\end{align}
for all $\ell \geq \ell_{hd}$, where we use $\ell_{hd}=100$, $\beta=1.5$, and $D=100$ and our simulation resolves spherical harmonic degrees until $\ell_{\rm{max}}=170$. We use 512 equidistant gridpoints in longitude, 256 equidistant gridpoints in latitude and 73 radial gridpoints on a Chebychev grid in all our simulations.

\begin{figure*}
    
    \centering
    \includegraphics[width=0.98\linewidth]{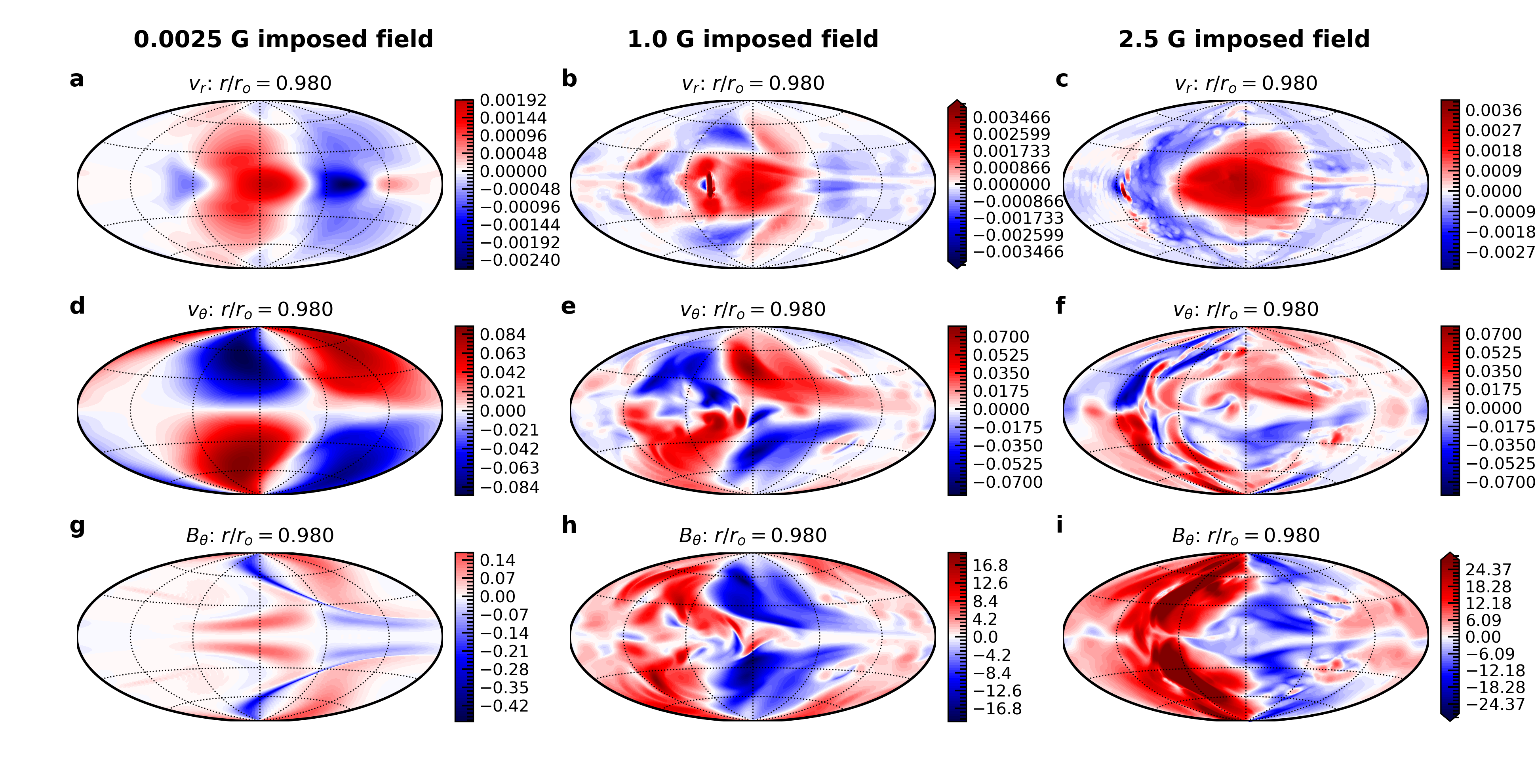}
    \caption{Radial flow (a-c), southward flow (d-f), and southward magnetic field (g-i) for the same runs and at the same depths as in Figure~\ref{figSubcritical}. Panel i is saturated at half of the maximum value attained in the displayed layer. Velocity is displayed in terms of the equatorial rotation speed and magnetic field is displayed in Gauss.}
    \label{figUrUThetaBtheta}
\end{figure*}

\begin{figure*}
    
    \centering
    \includegraphics[width=0.98\linewidth]{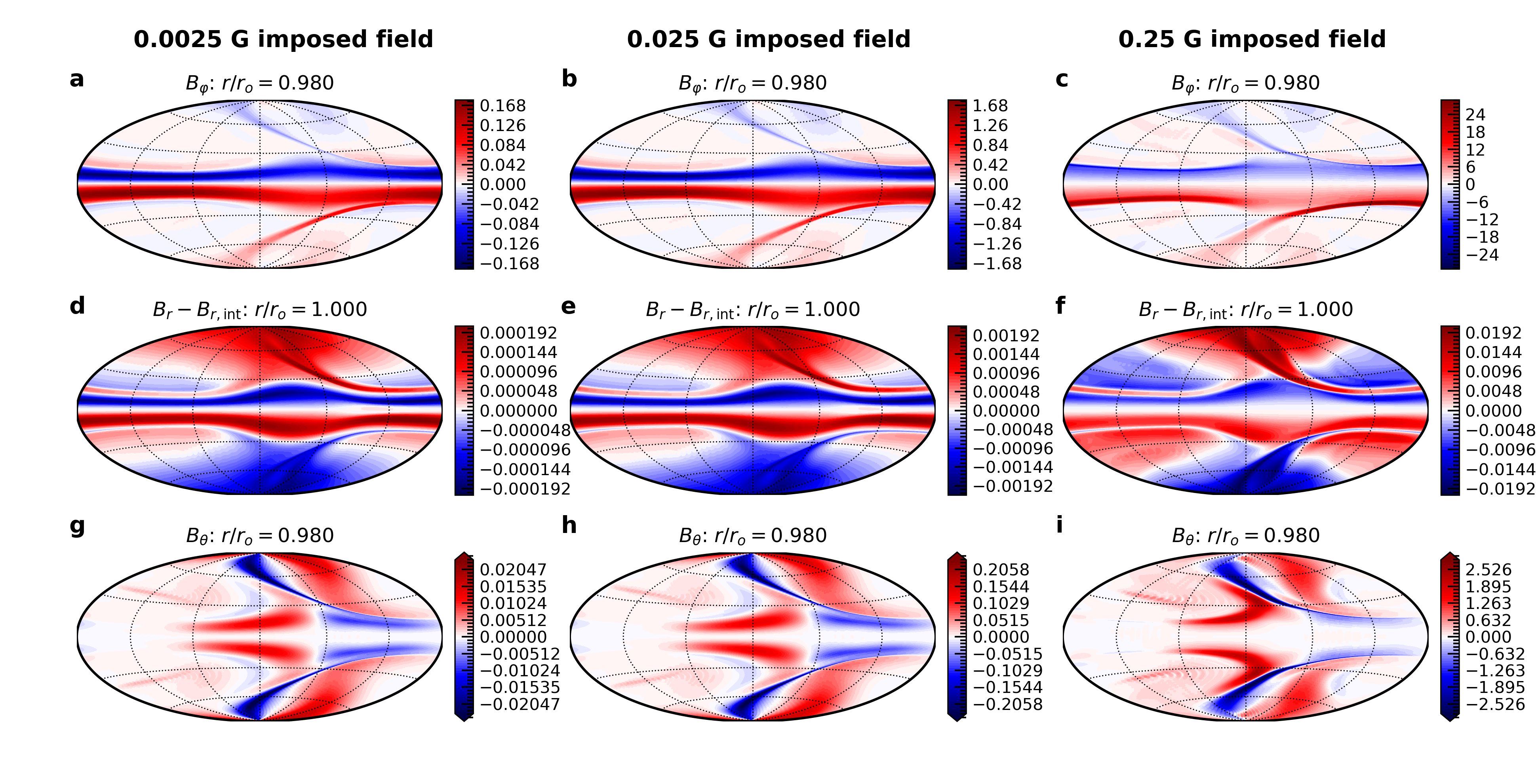}
    \caption{Azimuthal (a-c), radial (d-f), and southward (g-i) magnetic field for a three test runs with small imposed background magnetic field. Panels g-i are saturated at half of the maximum value attained in the displayed layer and panel. Magnetic field is displayed in Gauss.}
    \label{figWeakCases}
\end{figure*}

For the magnetic cases, we show results for the fields $u_r,u_\theta,B_\theta,B_{r,{\rm int}}$ in Figure~\ref{figUrUThetaBtheta}, which were not shown in Figure~\ref{figSubcritical}. Furthermore, in Figure~\ref{figWeakCases}, we show the results for a few test runs with weak magnetic fields, which were not fully discussed in this paper, but which complete the picture of transition from the linear to the non-linear regime.

\section{Energy and force spectra}
\label{appTimeEvolEnergiesSpectra}

\begin{figure*}
    \centering

    \includegraphics[width=0.98\linewidth]{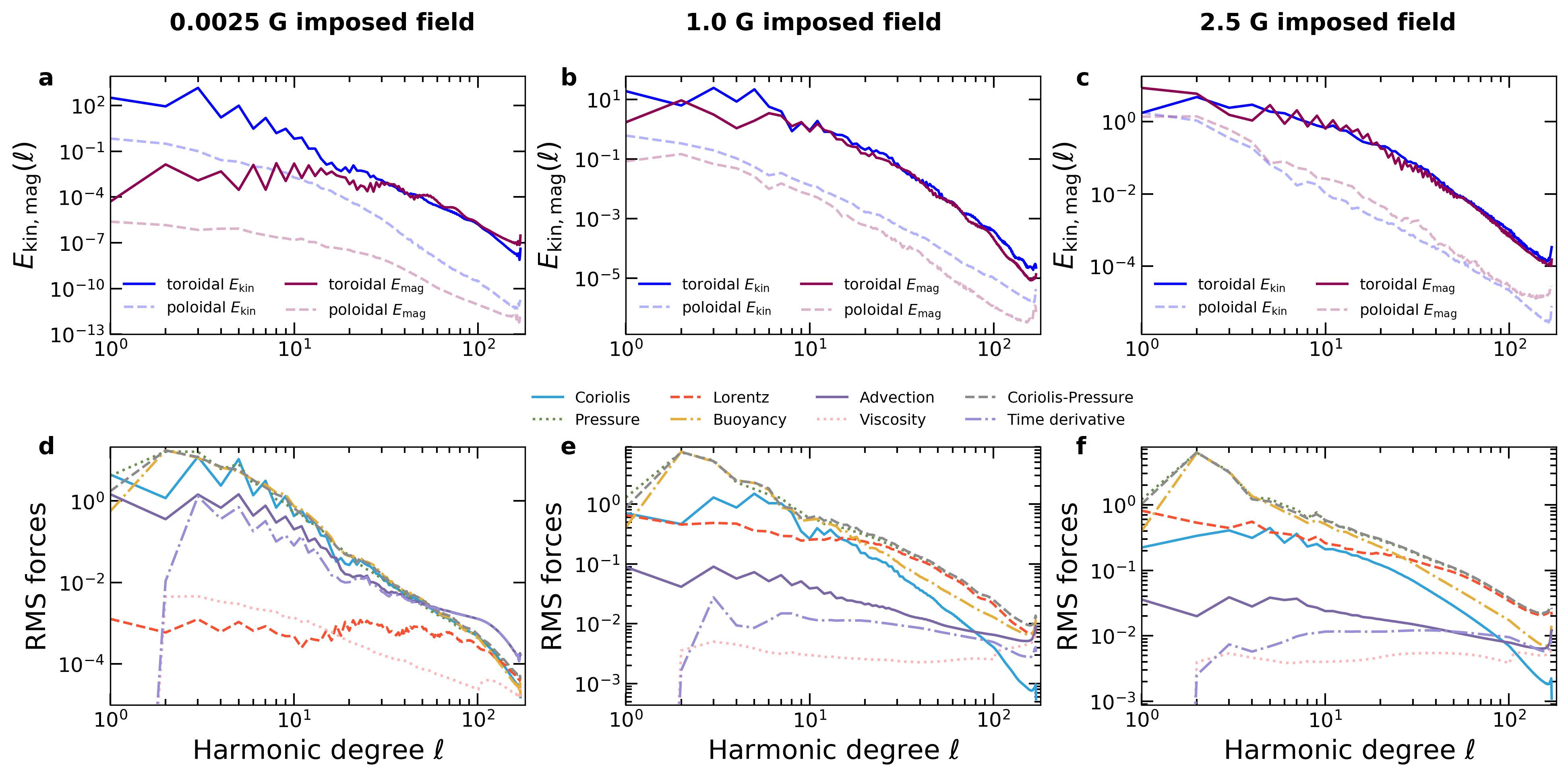}

    \caption{Magnetic and kinetic energy spectra (a-c) and rms. force spectra (d-f,) for the 0.025 G (a,d), 1.0 G (b,e), and 2.5 G (c,f) background dipolar field runs. The spectra of the dynamo run are very similar to the 2.5 G imposed field case.}
    \label{figSpectra}
\end{figure*}

In Figure~\ref{figSpectra}, we compare the kinetic and magnetic energy spectra (top row) and rms. force spectra \citep[bottom row,][]{Aubert2017,Schwaiger2019} for the three different background field amplitudes. The spectra for the dynamo run are by eye hard to distinguish from the spectra of the 2.5 G background field case. We find that only in the 2.5 G and dynamo runs, the magnetic energy spectrum exceeds the kinetic energy spectrum at large scales. In the other cases, the largest values are attained by the kinetic energy at large scales. 
The rms. force spectra show that the amplitude of the Lorentz force increases substantially with increasing forcing amplitude. Only in the 2.5 G and dynamo case, the Lorentz force exceeds the Coriolis force and has a dominant role in the force balance.

\section{Model for lightcurve, hotspot offset, and day-nightside temperature difference}

\label{appLightcurveHotspotShiftModel}

\begin{figure*}
    \centering

    \includegraphics[width=0.32\linewidth]{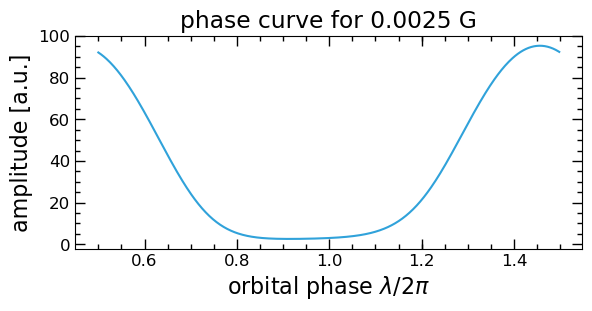}
    \includegraphics[width=0.32\linewidth]{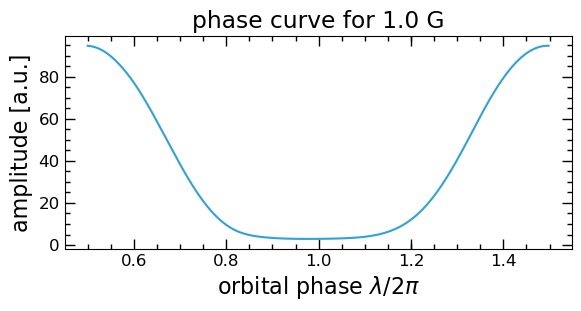}
    \includegraphics[width=0.32\linewidth]{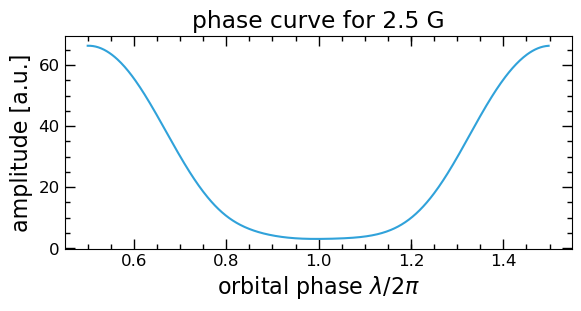}
    
    \caption{Total disc- and wave-length integrated intensity as a phasecurve model at $r_{\rm em}=0.98 r_o$. See Appendix~\ref{appLightcurveHotspotShiftModel} for details.}
    \label{figPhasecurves}
\end{figure*}

\begin{figure*}
    \centering

    \includegraphics[width=0.32\linewidth]{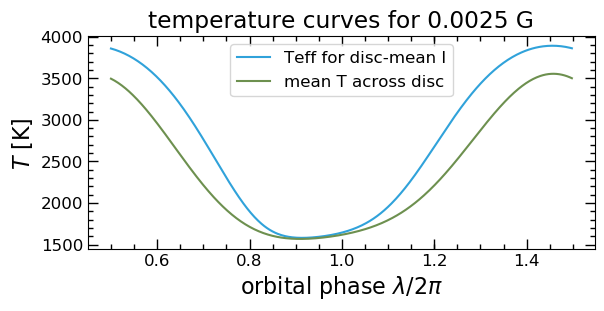}
    \includegraphics[width=0.32\linewidth]{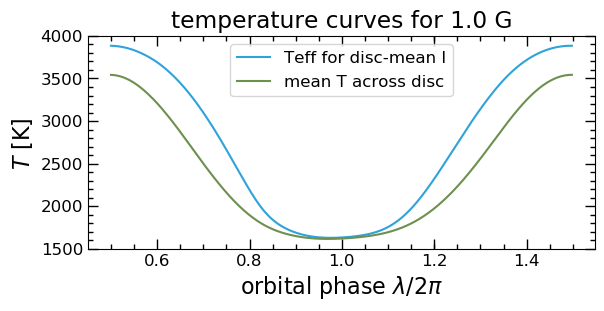}
    \includegraphics[width=0.32\linewidth]{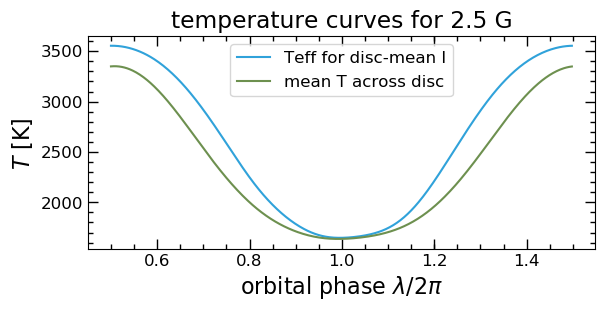}
    
    \caption{Disc-averaged effective and mean temperature as a function of orbital phase at $r_{\rm em}=0.98 r_o$. See Appendix~\ref{appLightcurveHotspotShiftModel} for details.}
    \label{figTempcurves}
\end{figure*}

To obtain a model for the lightcurves and to estimate the day-to-nightside temperature difference, we first need to convert from entropy to temperature. Using elementary thermodynamics and neglecting the impact of pressure changes, we find for a small volume $\Delta V$ with mass $\Delta m = \rho \Delta V$:
\begin{align}
    \id S &= \frac{\id Q}{T} \\
    \id s &= \frac{\id S}{\Delta V} \\
    \id Q 
    &= C_P \, \id T + \textrm{neglected pressure term} \\
    &= \Delta m \, c_P \, \id T\\
    &= \rho \,\Delta V \,c_P \,\id T \\
\Rightarrow    \id s &= \frac{\id S}{\Delta V} = \frac{\id Q}{T \, \Delta V} = \rho c_P \frac{\id T}{T}.
\end{align}
Assuming small values, we get from here for the dimensional quantities,
\begin{align}
   T' &= \frac{T}{\rho c_p} s' 
  \end{align}
and for the non-dimensional ones,
\begin{align}
  T' &= \tilde{T} \frac{s'}{\tilde \rho} .   
  \end{align}

This allows us to compute a model lightcurve from the entropy perturbations, as well as a hotspot offset and to estimate the day-nightside temperature difference. Since we do not have a proper radiative transfer model of the atmosphere yet coupled to our code, we propose to do this ad hoc, independently for every radial level in the atmosphere. For every level, we thereby effectively assume that it was the only radial level contributing to the emissions. Later, we will then choose a characteristic radial level as a phasecurve model. Since a wavelength-dependent model is currently not available to us, we propose to do this in a first step using the Stefan Boltzmann law, to at least obtain an estimate of what a phasecurve could look like in our case.

We compute the local irradiance using
\begin{align}
    I(r,\theta,\varphi,t)
    &= \sigma_{\rm SB} (\tilde{T}(r) + T'(r,\theta,\varphi))^4.
\end{align}
We model the visible light intensity at a certain CCD coordinate $(x,y)$, assumed radial emission level $r_{\rm em}$, and at a certain orbital phase $\lambda$ using 
\begin{align}
    x &= \sin \varphi' \sin \theta \\
    \varphi &= \pi - \lambda + \varphi' \\
    y &= \cos \theta \\
    dx &= \cos \varphi' \sin\theta \,\id\varphi' \\
    dy &= \sin\theta \,\id\theta.
\end{align}
This gives for the light curve
\begin{align}
    I(\lambda,r_{\rm em}) &= \int_{-1}^{1} \int_{-1}^{1} I(r_{\rm em},x,y) \,\id x \,\id y \\
    &= \iint I(r_{\rm em},\theta,\varphi)\,\cos \varphi'  \,\id\varphi' \sin^2\theta \,\id\theta,
\end{align}
where we integrate over the domains $\theta \in [0,\pi)$ and $\varphi'\in [-\pi/2,\pi/2]$.

The resulting phase curves $I(\lambda,r_{\rm em})$ are very smooth for all cases and all $r_{\rm em}$. From these curves, we extract the hotspot offset as
\begin{align}
    \lambda_{\rm hotspot}(r_{\rm em}) = \arg\max_{\lambda} I(\lambda,r_{\rm em}).
\end{align}

The dayside and nightside equilibrium temperatures are obtained from $I(\lambda,r_{\rm em})$ as
\begin{align}
    T_{\rm eff,day} (r_{\rm em})&= \left( \frac{I(\pi,r_{\rm em})}{\sigma_{\rm SB} \pi} \right)^{1/4}, \\
    T_{\rm eff,night} (r_{\rm em})&= \left( \frac{I(2\pi,r_{\rm em})}{\sigma_{\rm SB} \pi}\right)^{1/4},
\end{align}
from which we obtain the day-to-nightside temperature difference as
\begin{align}
    \Delta T(r_{\rm em}) &= T_{\rm eff,day} (r_{\rm em}) - T_{\rm eff,night} (r_{\rm em}).
\end{align}

Similarly, we obtain a disc-averaged mean temperature for comparison purposes as
\begin{align}
    \langle T\rangle_{\rm hem}(\lambda,r_{\rm em}) &= \frac{1}{\pi} \int_{-1}^{1} \int_{-1}^{1} T(r_{\rm em},x,y) \,\id x \,\id y \\
    &= \frac{1}{\pi} \iint T(r_{\rm em},\theta,\pi - \lambda + \varphi') \,\cos \varphi' \,\id\varphi' \sin^2\theta \,\id\theta,
\end{align}
where again $T = \tilde{T} + T'$.

Finally, we estimate the formation depth of the lightcurve in our model to be somewhat above the layer $r_{\rm eq}=0.954 r_o$, where $\tau_{\rm rad}=\tau_\Omega$ (see Sec.~\ref{appModel}), to be between $0.96 r_o \lesssim r_{\rm em} \lesssim 0.99 r_o$ by comparing our rotation and radiative timescales to the simulations of \citet{May2021}. Without further available information, we pick an intermediate value at $r_{\rm em}=0.98 r_o$ for illustration purposes in this paper. We show results for the resulting phase and temperature curves in Figures~\ref{figPhasecurves} and~\ref{figTempcurves}. The dependence of the resulting hotspot offset and day-to-nightside temperature difference on the assumed $r_{em}$ is shown in Figure~\ref{figHotspotshift}, where the possible range of $r_{em}$ is also indicated.

\end{document}